\documentclass{article}

\def\arcsec{\hbox{$^{\prime\prime}$}}
\def\arcmin{\hbox{$^\prime$}}
\def\degree{\hbox{$^{\rm o}$}}

\usepackage{bib}

\begin{document}
\title{$B$, $V$, $R$, $I$, $H$ and $K$ images of 86 face-on spiral galaxies}
\author{Roelof S. de Jong(1,2)\\
\ \\
\small (1) University of Durham, Dept.\ of Physics, South Road,
\small Durham DH1 3LE, United Kingdom\\
\small [R.S.deJong@durham.ac.uk]\\
\small (2) Kapteyn Astronomical Institute, P.O.box 800, 9700 AV
Groningen, The Netherlands
}

\maketitle

\begin{abstract}

FITS images in the $B$, $V$, $R$, $I$, $H$ and $K$ passbands are
presented of a sample of 86 face-on spiral galaxies.  The galaxies were
selected from the UGC to have a diameter of at least 2\arcmin\ and a
minor over major axis ratio larger than 0.625.  The selected galaxies
have an absolute Galactic latitude $|b| > 25$\degree, to minimize
the effect of Galactic extinction and foreground stars. 

Nearly all {\it BVRI\,} data were obtained with the 1m Jacobus Kapteyn
Telescope at La Palma and the $H$ and $K$ data were obtained at the 3.8m
UK Infra-Red Telescope at Hawaii.  The field of view of the
telescope/camera combinations were often smaller than the observed
galaxies, therefore driftscanning and mosaicing techniques were employed
to image at least along the major axis of the galaxies.  Most images were
obtained during photometric nights and calibrated using standard stars. 
A small fraction of the images was calibrated from literature aperture
photometry. 

The azimuthally averaged radial luminosity profiles derived from these
galaxy images (see de Jong and van der Kruit \cite{deJ1}, Paper~I) are
also made available in machine readable format, as are the results of
the bulge/disk decompositions described in de Jong (\cite{deJ2},
Paper~II).  A detailed statistical analysis of the bulge and disk
parameters of this data set can be found in de Jong (\cite{deJ3},
Paper~III).  The dust and stellar content of the galaxies as derived
from the color profiles is described in de Jong (\cite{deJ4}, Paper~IV). 
Evidence for secular evolution as found in this sample is shown in
Courteau, de Jong and Broeils (\cite{Cou96}). 


\end{abstract}

\section{Introduction}

A great deal about galaxy evolution can be learned by studying their
broadband properties.  Broadband observations give an immediate
impression of the spectral energy distribution and thereby information
on stellar and dust content.  Even though integrated magnitudes of
galaxies can be used to study global properties of galaxies, even more
can be learned from examining the detailed distribution of their light
and colors.  The star formation history in galaxies seems to be related
to their surface density properties (Kennicutt~\cite{Ken89}; Ryder and
Dopita~\cite{RydDop94}; de Jong~\cite{deJ4}), and therefore it is
imperative to have a statistical knowledge of surface brightness
distributions in galaxies to understand galaxy evolution. 

The image data set presented here was collected to study the surface
brightness distribution of spiral galaxies.  Of especial interest was
the question whether disks in spiral galaxies have a preferred central
surface brightness value as proposed by Freeman~(\cite{Fre70}).  The
observations were made in such a way that they were suitable to study
this central surface brightness effect, but this might make the
observations less useful for some other studies due to two limitations. 
(1) Disk central surface brightnesses are in general determined from
one-dimensional (1D) luminosity profiles, constructed by some kind of
azimuthal averaging of the light distribution.  No effort was made to
obtain images with high signal-to-noise per pixel, as large numbers of
pixels were to be averaged in the process of creating luminosity
profiles.  Furthermore the ``depth" of the optical images were matched
to the near-IR observations, which are more limited by the high sky
background level than by signal-to-noise ratios.  A considerable
fraction of the images have too low signal-to-noise per pixel to allow
detailed morphological studies of non-axisymmetric structures (ie.\ bars
and spiral arms) except in the highest surface brightness regions.  (2)
The used telescope/camera combinations had a limited field-of-view,
especially in the near-IR.  Often only the major axis was imaged of the
larger galaxies, as this was sufficient to measure the radial luminosity
distribution of the galaxy.  This again limits the usefulness of the
images to study non-axisymmetric light distributions in the outer part
of galaxies. 

The structure of this paper is as follows: the selection of the sample
is described in Section~2 and the observations in Section~3.  Section~4
explains the different data reduction techniques used.  In Section~5 I
describe the format of the FITS images on the CD-ROM, in Section~6 the
format of the luminosity profiles and in Section~7 the format of the
bulge/disk decomposition files.  A more detailed description of the
selection, observations and data reduction can be found in Paper~I.  The
bulge/disk decomposition methods are explained in more detail in
Paper~II.

\section{Selection}

The galaxies were selected from the Uppsala General Catalogue of
Galaxies (UGC, Nilson~\cite{UGC}).  Only spiral galaxies in the range
S1-DWARF\,SP were selected, excluding galaxies with classifications
as S0-S1, SB0-SB1, S3-IRR, IRR and DWARF IRR.  Ideally one would like
to have a volume-limited sample of galaxies for a statistical study of
galaxy properties, but this is impossible due to selection effects.  To
create a sample that is correctable for selection effects, the galaxies
were selected to have UGC red diameters of at least 2\arcmin.  The
galaxies have red UGC minor over major axis ratios larger than 0.625 to
reduce problems with projection effects and dust extinction.  This axis
ratio range corresponds to inclinations less than approximately 51\degree. 
Only galaxies with an absolute Galactic latitude $|b| > 25$\degree\ were
selected, to minimize the effect of Galactic extinction and to reduce
the number of foreground stars.  These selection criteria resulted in a
sample of 368 galaxies.  The final sample of 86 galaxies observed was
selected on the basis of hour angle and declination only, in such a way
that we had about equal number of observable galaxies during the whole
night in the granted observing time. The total selected areas cover about
12.5\% of the sky. All global parameters of the observed galaxies are
listed in Table~\ref{globpar}.

\section{Observations}

Nearly all {\it BVRI\,} images were obtained with the 1m Jacobus Kapteyn
Telescope (JKT) at La Palma, equipped with a 385x578 GEC CCD camera, in
March and September 1991 and April 1992.  The Kitt Peak {\it BVRI\,}
filter set (RGO / La Palma Technical Notes~\cite{RGO}) was used, the pixel
size was 0.3\arcsec.  The CCD camera was used in both its normal imaging
mode as well as in its driftscan mode.  In driftscan mode, optimal use is
made of the way CCDs are designed: while the telescope is tracking the
object, the CCD camera is shifted under the telescope at the same speed
as the image is shifted down the columns of the CCD while it is read
out.  Typical exposure times were 600\,s in $B$ and 400\,s for the other
optical passbands.  Twilight flatfields were obtained at the beginning
or at the end of the night and globular cluster fields with standard
stars were observed at regular intervals through the night for
calibration.  A small number of optical observations were obtained from
the La Palma archive. 

The near-IR $H$ and $K$ passband observations were made at the United
Kingdom Infrared Telescope at Hawaii with IRCAM~II containing a 58x62
InSb array.  During the February 1992 run standard $H$ and $K$ filters
were used, but a $K^\prime$ filter was used in September 1991.  The
pixel size was 1.2\arcsec.  For accurate sky subtraction and
flatfielding sky frames were obtained before and after every two object
frames at a position offset a few arcmin from the object.  Images were
taken in a strip along the major axis of the galaxies, spending about
twice as much time on the outer part of galaxies than on the central
region to increase signal-to-noise in these low surface brightness
regions.  Calibration stars from the list of Elias et al.~(\cite{Eli82})
were imaged at regular intervals.  Dark frames with exposure times equal
to the object exposure times were also obtained at regular intervals. 

The full observing log with observing method (driftscan, mosaic),
exposure times, photometric quality and seeing estimates can be found in
Paper~I.  These values are also store in the FITS headers of the images. 

\section{Data reduction}
\subsection{Optical data}

The normal data reduction procedure for CCD data was followed to create
calibrated images from the direct imaging data obtained with the JKT.  A
bias value was subtracted from the images using the average value in the
overscan region.  The images were divided by normalized flatfields
created by averaging several twilight frames.  No dark current was
subtracted as this was found to be insignificant for this CCD.  In
general two observations at the same position of an object were made,
which allowed cosmic-ray removal when they were averaged. 

The data reduction of the driftscans was more elaborate.  A driftscan
image consists of a ramp up part (rows that were not exposed for a full
chip length before being read out), a flat fully-exposed part and a ramp
down part (rows that are read out after the shutter has closed). 
The first rows of the ramp up part showed a gradient in the
bias level in the cross-scan direction.  Therefore, the bias level was
determined by fitting the first half of the ramp up part of each column,
giving a bias level for each column at the first row.  The images were
flatfielded by flatlines created averaging normal flatfields in column
direction. The ramp up and down parts were corrected for the shorter
exposure times, extending the field-of-view beyond the area that was
exposed to the sky for a full chip length.

\subsection{Near-IR data}

Careful attention had to be given to flatfielding of the near-IR images,
as flux levels 5$\times 10^4$ times below the sky level were measured. 
We first subtracted the dark current from all near-IR images (object and
sky) using the average of the two dark frames obtained nearest in time. 
A normalized flatfield image was created for each galaxy by taking the
median of the 4-5 sky frames observed around the galaxy.  After
flatfielding, known ``hot" and ``dead" pixels were set to ``undefined"
by a bad pixel mask and remaining dubious pixels were set to
``undefined' by hand.  These ``undefined" pixels were not used in
further analysis. 

The different object frames of a galaxy were mosaiced together to create
a full image along the major axis.  The spatial offset between frames
was determined by a cross-correlation technique or by using the
telescope offsets if no structure was available to be used in the
cross-correlation technique.  The relative spatial offsets between all
overlapping frame combinations were determined and a least-square-fit
determined the relative offset of all frames with respect to the central
frame.  Zero point (due to sky fluctuations) and intensity scaling
factors (only for non-photometric observations) were determined in a similar
fashion.  All zero point (and when necessary intensity) offsets between
overlapping frames (using the just determined spatial offsets) were
calculated and a least-squares-fit through all relative offsets provided
the intensity offset with respect to the central frame.  All frames were
mosaiced together using these spatial and intensity offsets, taking the
average in the overlapping areas. 

\subsection{Calibration}

The images were calibrated using the standard star fields observed
during each night under different airmasses.  The optical standard star
fields we used were calibrated to Landolt (\cite{Lan83}) stars, and
therefore our system response has been transformed to Johnson $B$ and
$V$ and Kron-Cousins $R$ and $I$.  The near-IR was calibrated to $H$ and
$K$ standard stars of Elias et al.~(\cite{Eli82}), using the corrections
of Wainscoat and Cowie~(\cite{WaiCow92}) to transform the $K^\prime$
passband to the $K$ passband.  Instrumental magnitudes (-2.5\,log(number
of counts)\,) of the different stars in the calibration fields were
measured with DAOphot (Stetson~\cite{Ste87}).  All photometric
calibration measurements of one observing run were combined to
least-square-fit equations of the form:
\begin{eqnarray} 
b & = & B + c_{0,B} + c_{1,B}(B-V) + c_{2,B} X \nonumber \\
v & = & V + c_{0,V} + c_{1,V}(B-V) + c_{2,V} X \nonumber \\
r & = & R + c_{0,R} + c_{1,R}(V-R) + c_{2,R} X \nonumber \\
i & = & I + c_{0,I} + c_{1,I}(R-I) + c_{2,I} X \nonumber \\
h & = & H + c_{0,H} + c_{2,H} X \nonumber \\
k & = & K + c_{0,K} + c_{2,K} X
\label{caleq}
\end{eqnarray}
 where $B, V, R$, $I$, $H$ and $K$ are the standard star magnitudes, $b,
v, r$, $i$, $h$ and $k$ the instrumental magnitudes per second, $X$ the
airmass of the observation and $c_{i,J}$ the unknown transformation
coefficients.  The results of these fits can be found in
Tables~\ref{mag0tab} and \ref{mag0tabir} and in the FITS headers of the
images.  

Non-photometric observations were calibrated with aperture photometry
from the literature when available.  We first determined magnitudes in
synthetic apertures of the size of the literature photometry using the
calibration of a photometric night.  If our magnitude differed more than
the expected error from the literature value, all magnitude parameters
were corrected for this difference (indicated by header item CORR in the
FITS files).

The optical pixel size was determined to be 0.303$\pm$0.004\arcsec, using
images of globular clusters which contained accurately known star
positions.  This pixel size agreed to within its uncertainty to the
instrumental specification, and therefore a value of 0.30\arcsec\ was
adopted.  The near-IR pixel size was derived from the scaling factor to
align the near-IR images with the optical images (see next paragraph). 
The near-IR pixel size was 1.20\arcsec\ per pixel. 

\subsection{Final reduction steps}

We determined the sky background level on the fully reduced images using
the box method.  Average sky values were measured in small boxes around
the galaxies.  Sky level was set to the median value of these
measurements.  The uncertainty in the sky value was taken to be half the
difference between the maximum and minimum average sky values found in
these boxes. This uncertainty will reflect errors due to imperfect
flatfielding and mosaicing.

We aligned the images in the different passbands using foreground stars
in common between the different frames.  Images obtained during the same
observing run were only allowed to shift, between different runs also
rotation and scaling was allowed.  The near-IR data was regridded to the
much smaller pixel scale of the optical images, which means that nothing
smaller than the original pixel size (1.2\arcsec) should be trusted on
these images.  A linear interpolation was used for regridding and
therefore the new smaller pixels contain values that are representative
of the original surface brightness in the pixels of the original size. 
Total flux in the image is not conserved in this process, but the
original number of counts in an area can easily be calculated by
multiplying the new number of counts in an area with the ratio of the
square of the pixel sizes, (pixelsize$_{\rm new}$/pixelsize$_{\rm
old}$)$^2$.

\section{The image catalog}

All aligned images are stored in FITS format on the CD-ROM in the
directory {\tt images/}, with a separate directory for each galaxy.  The
aligned near-IR images in these directories have been compressed with
gzip, but the ``raw'' near-IR images (ie.\ before aligning and
regridding to the optical images) are available in uncompressed FITS
format in the directory {\tt IRimages/}.  The FITS headers contain all
the essential information for analysis.  The images are in
analog-to-digital-units (ADU), which corresponds approximately to the
number of detected photons for the optical images and to 50 detected
photons in the near-IR images.  Undefined pixels in the images contain
the value -999.  The header items of interest are as follows:

\smallskip
\noindent
{\underline{\bf Basic FITS items}}
\begin{description}
\item[NAXIS1, NAXIS2] number of pixels in RA and DEC respectively
\item[CTYPE1, CTYPE2] RA-TAN, DEC-TAN axis type and projection system
\item[CRVAL1, CRVAL2] should contain the RA and DEC value at the reference 
pixel ({\bf CRPIX1, CRPIX2}), but as the exact position of the galaxies 
was often unknown, the stored values have no meaning
\item[CDELT1, CDELT2] the pixel size in {\it degrees}. The same value is
stored in arcseconds in header item {\bf PIXSIZIM}
\end{description}

\noindent
{\underline{\bf Observation related}}
\begin{description}
\item[FILTER] passband filter (B, V, R, I, H, K or K$^\prime$)
\item[SEEING] full-width-at-half-maximum (FWHM) of seeing estimate in arcsec
\item[PHOT] photometric quality estimate as in Paper~I (1:
photometric, 2: 0.0-0.2 mag, 3: 0.2-0.5 mag, 4: 0.5-1.0 mag and 5:
$>$1.0 mag error)
\item[QUAL] quick look quality estimate, taking into account (in order
of importance) flatfield quality, area to measure the sky level,
signal-to-noise and seeing. The numbers mean, 1: excellent, 2: reasonable, 
but take into account some of the limitations such as limited sky area, 
3: poor, do not use except in case of an emergency
\end{description}

\noindent
{\underline{\bf Calibration}}
\begin{description}
\item[MAG0] zero point calibration constant for a 1 second exposure
(-$c_0$ in Eq.~\ref{caleq})
\item[CCOL] color calibration constant, when not used 0 ($c_1$)
\item[COL] average color of this galaxy used for calibration
\item[CAIR] airmass calibration constant ($c_2$)
\item[AIRMASS] airmass during the observation ($X$)
\item[CORR] correction for non-photometric observation to put this image
on literature photometry
\item[PIXSIZE] pixel size in arcsec of original image (before
rebinning)
\item[PIXSIZIM] pixel size in arcsec of this image (after
rebinning/aligning)
\item[EXPTIME] exposure time calibration constant (if several images
were averaged, this contains the average exposure time)
\item[SKYLEV] estimate of the sky background level in ADU
\item[SKYERR] maximum uncertainty in sky background
 \item[MAGOFF] for convenience, this constant gives the calibration to
convert pixel ADU values into mag arcsec$^{-2}$.  It is equal to {\bf
$-$MAG0$-$CCOL$\times$COL$-$CAIR$\times$AIRMASS $-$ CORR
$+2.5\log($PIXSIZE$^2$$\times$EXPTIME}).  The surface brightness in mag
arcsec$^{-2}$ of a pixel with ADU counts in the galaxy is {\bf
MAGOFF}$-2.5\log($pixel$($ADU$)-${\bf SKYLEV}$)$.  To use this constant
to calculate the magitude in an area, take into acount that flux was not
conserved per area in the rebinning/alligning proces.  The magnitude in
an area with total of ADU counts is {\bf
MAGOFF$-2.5\log($}area$($ADU$)-${\bf
SKYLEV$)-2.5\log($PIXSIZE$^4/$PIXSIZIM$^2)$}

\end{description}

\section{Luminosity profiles}

The radial luminosity distribution of each galaxy was determined in each
passband and these are also present on the CD-ROM.  The areas in the $R$
passband images affected by foreground stars were masked using a polygon
editor.  This mask was transfered to the other passbands, thus making
certain that the same area was used in all passbands.  The center of the
galaxy was determined by fitting an ellipse to the central peak in the
$R$ passband image.  Next, with this center fixed, ellipses were fit to
the isophotes at the 23.5, 24.0 and 24.5 $R$-mag arcsec$^{-2}$ level. 
The median values found for the minor/major axis ratio ($b/a$) and
position angle (PA) in the $R$-band were used in all passbands to
determine the luminosity profiles.  Average ADU values were determined
in concentric elliptical annuli of increasing radius with the already
determined center, $b/a$ and PA fixed.  For face-on galaxies this method
gives a better estimate of the average luminostity at each radius than
methods which freely fit ellipses at each isophote, if we assume that
the galaxy is not strongly warped. Bars, spiral arms and H{\sc II}
regions make isophote fitting methods unreliable for face-on spiral galaxies.

The profiles are provided in ASCII in the directory {\tt Profiles/} and
the graphs can be found in Paper~I.  The surface brightness profiles are
in mag arcsec$^{-2}$, the radii in arcsec.  Undefined values are
indicated by a \raisebox{-0.8mm}{*}.  Note that the central regions of
UGC\,7540 were saturated in the $V$, $R$ and $I$ passband.  Further
header information in these files are

\begin{description}
\item[INCL] inclination in degrees (actually cos$^{-1}$($b/a$)) used for 
profile extraction
\item[PA] position angle in degrees used for profile extraction,
measured from north to east
\item[EXPTIME] exposure time in seconds of image used
\item[MAGOFF] magnitude calibration constant (see image catalog)
\item[MAGSKY] sky surface brightness in mag arcsec$^{-2}$
\item[MAGSKYERR] uncertainty in {\bf MAGSKY}
\item[MAGTOT] total apparent magnitude of the galaxy derived from the surface
brightness profile (see Paper~I)
\item[MAGERR] uncertainty in apparent magnitude
\item[SEEING] FWHM of seeing estimate in arcsec
\item[DATE] date of observation
\item[PHOTQ] photometry quality estimate (see image catalog)
\end{description}

\section{Bulge/disk decompositions}

A number bulge/disk decomposition methods was applied to the data (see
Paper~II for details) and the results are stored in directory {\tt
B\_\,Dratio/}.  The results of the 1D profile decompositions with
$R^{1/4}$, $R^{1/2}$ and exponential bulges can be found in the files
{\tt bd4qfpar.dat}, {\tt bd4ffpar.dat} and {\tt bd4efpar.dat}
respectively.  The results of the 2D decompositions with exponential
bulges and disks and with Freeman bars can be found in {\tt
bd4fpar.dat}.  Note that not all observations were photometric and that
for non-photometric observations the listed numbers are the lower limits
in surface brightness flux.  Obviously the scale parameters are correct
for the non-photometric observations.  Check the file {\tt pht.dat} for
a listing of the photometric quality of the observations.  The
description of all the columns in these files can be found in file {\tt
bd4Read.Me}.

\section*{Acknowledgements}

This research was supported under grant no.~782-373-044 from the
Netherlands Foundation for Research in Astronomy (ASTRON), which
receives its funds from the Netherlands Foundation for Scientific
Research (NWO).  This paper is based on observations with the Jacobus Kapteyn
Telescope and the Isaac Newton Telescope operated by the Royal
Greenwich Observatory at the Observatorio del Roque de los Muchachos of
the Instituto de Astrof\'\i sica de Canarias with financial support
from the PPARC (UK) and NWO (NL) and with the UK Infrared Telescope at
Mauna Kea operated by the Royal Observatory Edinburgh with financial
support of the PPARC. 


%
%

{
\tabcolsep=1.mm

\begin{table}
\caption
 {Global parameters of the galaxies in the observed sample.  The
positions and the $V_{\rm GSR}$ recession velocities (cz) are obtained
from the RC3 catalog, $D_{\rm maj}$ is the red UGC major axis diameter,
$b/a$ is the red UGC minor over major axis diameter ratio. 
 \label{globpar}
}
\begin{tabular}{llr@{$\ \,$}r@{$\ \,$}r@{$\ \ \ $}r@{$\ \,$}r@{$\ \,$}rccccr}
\ \\
\hline
\hline
\multicolumn{2}{c}{name}& \multicolumn{3}{c}{RA} &\multicolumn{3}{c}{DEC}&  \multicolumn{2}{c}{classification} & $D_{\rm maj}$ & $b/a$ & $V_{\rm GSR}$\\
\phantom{ UGC 12345}&  & \multicolumn{6}{c}{(1950)}      &UGC & RC3     & (\arcmin)& & km/s\\
\hline
 UGC    89 & NGC    23  &  0 & 07 & 18.6 & 25 &  38 & 42 & SB1     & .SBS1.. &  2.2 &  0.68 &  4733 \\
 UGC    93 &            &  0 & 07 & 47.0 & 30 &  34 & 16 & S IV    & .SA.8.. &  2.0 &  0.85 &  5124 \\
 UGC   242 &            &  0 & 22 & 52.6 & 19 &  57 & 39 & SB3     & .SX.7.. &  2.1 &  0.86 &  4449 \\
 UGC   334 & A   0031+31&  0 & 31 & 16.6 & 31 &  10 & 33 & DWRF SP & .S..9.. &  2.0 &  1.00 &  4800 \\
 UGC   438 & NGC   214  &  0 & 38 & 48.9 & 25 &  13 & 33 & S3      & .SXR5.. &  2.2 &  0.77 &  4685 \\
 UGC   463 & NGC   234  &  0 & 40 & 55.6 & 14 &  04 & 10 & S3      & .SXT5.. &  2.0 &  1.00 &  4577 \\
 UGC   490 & NGC   251  &  0 & 45 & 12.0 & 19 &  18 & 00 & S3      & .S..5.. &  2.3 &  0.78 &  4732 \\
 UGC   508 & NGC   266  &  0 & 47 & 05.6 & 32 &  00 & 23 & SB1     & .SBT2.. &  3.5 &  0.94 &  4823 \\
 UGC   628 &            &  0 & 58 & 18.0 & 19 &  13 & 00 & DWRF SP & .S..9*. &  2.0 &  0.80 &  5574 \\
 UGC  1305 & NGC   691  &  1 & 47 & 55.8 & 21 &  30 & 45 & S2/S3   & .SAT4.. &  3.7 &  0.70 &  2769 \\
 UGC  1455 & NGC   765  &  1 & 55 & 58.7 & 24 &  38 & 56 & SB2/S3  & .SXT4.. &  3.0 &  1.00 &  5224 \\
 UGC  1551 &            &  2 & 00 & 48.4 & 23 &  50 & 03 & SB IV-V & .SB?... &  3.0 &  0.67 &  2773 \\
 UGC  1559 & IC   1774  &  2 & 01 & 12.0 & 15 &  04 & 00 & S3/SB3  & .SXS7.. &  2.1 &  0.81 &  3705 \\
 UGC  1577 &            &  2 & 02 & 32.3 & 30 &  56 & 14 & SB2     & .SB?... &  2.3 &  0.70 &  5393 \\
 UGC  1719 & IC    213  &  2 & 11 & 18.0 & 16 &  14 & 00 & S2      & .SXT3.. &  2.2 &  0.73 &  8297 \\
 UGC  1792 &            &  2 & 16 & 58.2 & 28 &  48 & 27 & SB3     & .SXR5.. &  2.2 &  0.64 &  5092 \\
 UGC  2064 &            &  2 & 32 & 18.0 & 20 &  38 & 00 & SB2/S3  & .SXS4.. &  2.1 &  0.71 &  4338 \\
 UGC  2081 &            &  2 & 33 & 27.1 &  0 &  12 & 08 & S3      & .SXS6.. &  2.5 &  0.72 &  2626 \\
 UGC  2124 & NGC  1015  &  2 & 35 & 38.9 & -1 &  32 & 00 & SB1     & .SBR1*. &  3.0 &  1.00 &  2639 \\
 UGC  2125 & IC   1823  &  2 & 35 & 36.9 & 31 &  51 & 14 & SB3     & .SBR5.. &  2.3 &  0.87 &  5288 \\
 UGC  2197 &            &  2 & 40 & 25.8 & 31 &  15 & 34 & S3      & .S..6*. &  2.0 &  0.70 &  5195 \\
 UGC  2368 & IC    267  &  2 & 51 & 06.1 & 12 &  38 & 43 & SB2     & PSBS3.. &  2.1 &  0.71 &  3610 \\
 UGC  2595 & IC    302  &  3 & 10 & 13.9 &  4 &  31 & 06 & SB2/SB3 & .SBT4.. &  2.5 &  0.92 &  5907 \\
 UGC  3066 &            &  4 & 28 & 18.2 &  5 &  26 & 00 & S3/SB3  & .SXR7*. &  2.0 &  0.75 &  4594 \\
 UGC  3080 & A   0429+01&  4 & 29 & 21.8 &  1 &  05 & 27 & S3      & .SXT5.. &  2.2 &  1.00 &  3481 \\
 UGC  3140 & NGC  1642  &  4 & 40 & 20.1 &  0 &  31 & 35 & S3      & .SAT5*. &  2.0 &  1.00 &  4564 \\
 UGC  4126 & NGC  2487  &  7 & 55 & 19.0 & 25 &  17 & 08 & SB2     & .SB.3.. &  2.5 &  0.92 &  4771 \\
 UGC  4256 & NGC  2532  &  8 & 07 & 03.2 & 34 &  06 & 20 & S3      & .SXT5.. &  2.2 &  0.82 &  5228 \\
 UGC  4308 & A   0814+21&  8 & 14 & 29.9 & 21 &  50 & 20 & SB3     & .SBT5.. &  2.2 &  0.77 &  3486 \\
 UGC  4368 & NGC  2575  &  8 & 19 & 46.2 & 24 &  27 & 32 & S3      & .SAT6*. &  2.5 &  0.80 &  3800 \\
 UGC  4375 & A   0820+22&  8 & 20 & 12.0 & 22 &  49 & 00 & S3      & .SX.5*. &  2.5 &  0.68 &  1983 \\
 UGC  4422 & NGC  2595  &  8 & 24 & 46.7 & 21 &  38 & 40 & SB2/S3  & .SXT5.. &  3.2 &  0.88 &  4250 \\
 UGC  4458 & NGC  2599  &  8 & 29 & 15.4 & 22 &  44 & 00 & S1      & .SA.1.. &  2.0 &  1.00 &  4672 \\
 UGC  5103 & NGC  2916  &  9 & 32 & 07.6 & 21 &  55 & 45 & S       & .SAT3\$.&  2.3 &  0.74 &  3649 \\
 UGC  5303 & NGC  3041  &  9 & 50 & 22.5 & 16 &  54 & 53 & S3      & .SXT5.. &  3.8 &  0.63 &  1317 \\
 UGC  5510 & NGC  3162  & 10 & 10 & 45.5 & 22 &  59 & 16 & S3      & .SXT4.. &  3.2 &  0.88 &  1231 \\
 UGC  5554 & NGC  3185  & 10 & 14 & 53.2 & 21 &  56 & 20 & SB1     & RSBR1.. &  2.8 &  0.64 &  1159 \\
 UGC  5633 & A   1021+15& 10 & 21 & 54.0 & 15 &  00 & 00 & SB IV-V & .SB.8.. &  2.5 &  0.64 &  1287 \\
 UGC  5842 & NGC  3346  & 10 & 40 & 59.0 & 15 &  08 & 03 & SB3     & .SBT6.. &  3.0 &  0.87 &  1169 \\
 UGC  6028 & NGC  3455  & 10 & 51 & 51.6 & 17 &  33 & 08 & S2      & PSXT3.. &  2.6 &  0.65 &  1029 \\
 UGC  6077 & NGC  3485  & 10 & 57 & 24.0 & 15 &  06 & 43 & SB2     & .SBR3*. &  2.3 &  1.00 &  1350 \\
 UGC  6123 & NGC  3507  & 11 & 00 & 46.3 & 18 &  24 & 25 & SB2     & .SBS3.. &  3.4 &  0.82 &   906 \\
\hline
\hline
\end{tabular}
\end{table}

\addtocounter{table}{-1}
\begin{table}
\caption{-continued.}

\begin{tabular}{llr@{$\ \,$}r@{$\ \,$}r@{$\ \ \ $}r@{$\ \,$}r@{$\ \,$}rccccr}
\ \\
\hline
\hline
\multicolumn{2}{c}{name}& \multicolumn{3}{c}{RA} &\multicolumn{3}{c}{DEC}&  \multicolumn{2}{c}{classification} & $D_{\rm maj}$ & $b/a$ & $V_{\rm GSR}$\\
\phantom{ UGC 12345}&  & \multicolumn{6}{c}{(1950)}      & UGC & RC3     & (\arcmin)& & km/s\\
\hline
 UGC  6277 & NGC  3596  & 11 & 12 & 27.9 & 15 &  03 & 38 & S3      & .SXT5.. &  3.6 &  0.78 &  1111 \\
 UGC  6445 & NGC  3681  & 11 & 23 & 52.6 & 17 &  08 & 22 & S2/S3   & .SXR4.. &  2.3 &  1.00 &  1171 \\
 UGC  6453 & NGC  3684  & 11 & 24 & 34.4 & 17 &  18 & 20 & S3      & .SAT4.. &  2.5 &  0.68 &  1097 \\
 UGC  6460 & NGC  3686  & 11 & 25 & 07.3 & 17 &  29 & 56 & SB2/SB3 & .SBS4.. &  3.0 &  0.83 &  1089 \\
 UGC  6536 & NGC  3728  & 11 & 30 & 36.0 & 24 &  43 & 00 & S2      & .S..3.. &  2.0 &  0.75 &  6941 \\
 UGC  6693 & NGC  3832  & 11 & 40 & 54.0 & 23 &  00 & 00 & SB3     & .SBT4.. &  2.2 &  0.95 &  6869 \\
 UGC  6746 & NGC  3884  & 11 & 43 & 37.0 & 20 &  40 & 11 & S1      & .SAR0.. &  2.1 &  0.81 &  6897 \\
 UGC  6754 & NGC  3883  & 11 & 44 & 11.5 & 20 &  57 & 16 & S2      & .SAT3.. &  3.3 &  0.91 &  6979 \\
 UGC  7169 & NGC  4152  & 12 & 08 & 04.6 & 16 &  18 & 42 & S3      & .SXT5.. &  2.2 &  0.86 &  2112 \\
 UGC  7315 & NGC  4237  & 12 & 14 & 38.2 & 15 &  36 & 08 & S2      & .SXT4.. &  2.2 &  0.64 &   813 \\
 UGC  7450 & NGC  4321  & 12 & 20 & 23.3 & 16 &  06 & 00 & S3      & .SXS4.. &  6.8 &  0.88 &  1540 \\
 UGC  7523 & NGC  4394  & 12 & 23 & 24.7 & 18 &  29 & 30 & SB2     & RSBR3.. &  3.9 &  0.90 &   884 \\
 UGC  7594 & NGC  4450  & 12 & 25 & 58.0 & 17 &  21 & 40 & S2      & .SAS2.. &  6.5 &  0.69 &  1918 \\
 UGC  7876 & NGC  4635  & 12 & 40 & 09.5 & 20 &  13 & 12 & S3      & .SXS7.. &  2.0 &  0.80 &   938 \\
 UGC  7901 & NGC  4651  & 12 & 41 & 12.5 & 16 &  40 & 05 & S3      & .SAT5.. &  4.0 &  0.68 &   772 \\
 UGC  8279 & NGC  5016  & 13 & 09 & 42.6 & 24 &  21 & 42 & S2-3    & .SXT5.. &  2.0 &  0.75 &  2622 \\
 UGC  8289 & NGC  5020  & 13 & 10 & 11.0 & 12 &  51 & 53 & S2/SB3  & .SXT4.. &  3.3 &  0.85 &  3331 \\
 UGC  8865 & NGC  5375  & 13 & 54 & 40.7 & 29 &  24 & 26 & SB2     & .SBR2.. &  3.7 &  0.81 &  2435 \\
 UGC  9024 &            & 14 & 04 & 24.0 & 22 &  16 & 00 & S       & .S?.... &  2.0 &  1.00 &  2338 \\
 UGC  9061 & IC    983  & 14 & 07 & 42.4 & 17 &  58 & 08 & SB1/SB2 & .SBR4.. &  4.5 &  0.78 &  5466 \\
 UGC  9481 & NGC  5735  & 14 & 40 & 23.5 & 28 &  56 & 15 & SB2     & .SBT4.. &  2.2 &  0.82 &  3817 \\
 UGC  9915 & NGC  5957  & 15 & 33 & 00.9 & 12 &  12 & 51 & SB2     & PSXR3.. &  2.8 &  1.00 &  1889 \\
 UGC  9926 & NGC  5962  & 15 & 34 & 14.1 & 16 &  46 & 23 & S3      & .SAR5.. &  2.8 &  0.71 &  2034 \\
 UGC  9943 & NGC  5970  & 15 & 36 & 08.1 & 12 &  20 & 53 & SB3     & .SBR5.. &  2.9 &  0.66 &  2030 \\
 UGC 10083 & NGC  6012  & 15 & 51 & 54.6 & 14 &  44 & 55 & SB1     & RSBR2*. &  2.0 &  0.65 &  1944 \\
 UGC 10437 &            & 16 & 29 & 36.0 & 43 &  27 & 00 & S       & .S?.... &  2.0 &  0.85 &  2759 \\
 UGC 10445 &            & 16 & 31 & 48.6 & 29 &  05 & 19 & S3      & .S..6?. &  2.3 &  0.87 &  1102 \\
 UGC 10584 & NGC  6246A & 16 & 49 & 12.0 & 55 &  28 & 00 & S3/SB3  & .SXR5P* &  2.3 &  0.91 &  5451 \\
 UGC 11628 & NGC  6962  & 20 & 44 & 45.4 &  0 &  08 & 13 & S1      & .SXR2.. &  3.0 &  0.77 &  4370 \\
 UGC 11708 & NGC  7046  & 21 & 12 & 24.1 &  2 &  37 & 38 & SB      & .SBT6.. &  2.0 &  0.65 &  4326 \\
 UGC 11872 & NGC  7177  & 21 & 58 & 18.6 & 17 &  29 & 50 & S2      & .SXR3.. &  2.7 &  0.70 &  1343 \\
 UGC 12151 &            & 22 & 39 & 00.0 &  0 &  08 & 00 & DWARF   & .IBS9*. &  3.0 &  0.67 &  1896 \\
 UGC 12343 & NGC  7479  & 23 & 02 & 26.8 & 12 &  03 & 06 & SB2     & .SBS5.. &  4.0 &  0.83 &  2544 \\
 UGC 12379 & NGC  7490  & 23 & 05 & 01.0 & 32 &  06 & 18 & S2      & .S..4.. &  2.3 &  1.00 &  6416 \\
 UGC 12391 & NGC  7495  & 23 & 06 & 24.0 & 11 &  46 & 00 & S3      & .SXS5.. &  2.0 &  0.85 &  5050 \\
 UGC 12511 & NGC  7610  & 23 & 17 & 09.8 &  9 &  54 & 40 & S3      & .S..6*. &  2.5 &  0.84 &  3708 \\
 UGC 12614 & NGC  7678  & 23 & 25 & 58.2 & 22 &  08 & 50 & S3/SB3  & .SXT5.. &  2.8 &  0.68 &  3665 \\
 UGC 12638 & NGC  7685  & 23 & 28 & 00.2 &  3 &  37 & 31 & S3      & .SXS5*. &  2.0 &  0.85 &  5775 \\
 UGC 12654 & NGC  7691  & 23 & 29 & 53.0 & 15 &  34 & 28 & SB2/S3  & .SXT4.. &  2.0 &  0.80 &  4224 \\
 UGC 12732 &            & 23 & 38 & 09.1 & 25 &  57 & 30 & DWRF SP & .S..9*. &  3.0 &  1.00 &   929 \\
 UGC 12754 & NGC  7741  & 23 & 41 & 22.7 & 25 &  47 & 53 & SB3     & .SBS6.. &  4.3 &  0.70 &   935 \\
 UGC 12776 &            & 23 & 43 & 41.4 & 33 &  05 & 26 & SB2     & .SBT3.. &  2.7 &  0.81 &  5127 \\
 UGC 12808 & NGC  7769  & 23 & 48 & 31.5 & 19 &  52 & 25 & S1-2    & RSAT3.. &  2.5 &  0.84 &  4380 \\
 UGC 12845 &            & 23 & 53 & 11.0 & 31 &  37 & 23 & S3      & .S..7.. &  2.4 &  0.75 &  5064 \\
\hline
\hline
\end{tabular}
\end{table}
}


{
\tabcolsep=1.4mm
\begin{table}
\begin{center}
\caption[]{
Calibration coefficients determined for the different observing runs 
on the JKT.
}
\begin{tabular}{cccc}
\ \\
\hline
\hline
\ \ passband & \ zero-point ($c_0$)\ & \ \ color coef.~($c_1$)\ & extinction coef.~($c_2$)\\
\hline
\multicolumn{4}{c}{April 3-9, 1991}\\
$B$& -22.251\,$\pm$\,0.065 & -0.062\,$\pm$\,0.011 & 0.251\,$\pm$\,0.027\\
$V$& -22.791\,$\pm$\,0.032 & -0.013\,$\pm$\,0.007 & 0.216\,$\pm$\,0.030\\
$R$& -22.883\,$\pm$\,0.030 & -0.001\,$\pm$\,0.010 & 0.179\,$\pm$\,0.020\\
$I$& -22.060\,$\pm$\,0.045 & -0.012\,$\pm$\,0.015 & 0.058\,$\pm$\,0.058\\
\hline
\multicolumn{4}{c}{September 7-10, 1991}\\
$B$& -21.757\,$\pm$\,0.111 & -0.161\,$\pm$\,0.044 & 0.238\,$\pm$\,0.065\\
$V$& -22.215\,$\pm$\,0.067 & -0.048\,$\pm$\,0.024 & 0.135\,$\pm$\,0.025\\
$R$& -22.438\,$\pm$\,0.073 & -0.016\,$\pm$\,0.046 & 0.141\,$\pm$\,0.020\\
$I$& -21.709\,$\pm$\,0.081 & -0.034\,$\pm$\,0.057 & 0.081\,$\pm$\,0.082\\
\hline
\multicolumn{4}{c}{September 13-16, 1991}\\
$B$& -21.977\,$\pm$\,0.122 & -0.161\,$\pm$\,0.044 & 0.279\,$\pm$\,0.052\\
$V$& -22.322\,$\pm$\,0.072 & -0.048\,$\pm$\,0.024 & 0.121\,$\pm$\,0.030\\
$R$& -22.558\,$\pm$\,0.064 & -0.016\,$\pm$\,0.046 & 0.126\,$\pm$\,0.026\\
$I$& -21.833\,$\pm$\,0.068 & -0.034\,$\pm$\,0.057 & 0.023\,$\pm$\,0.027\\
\hline
\multicolumn{4}{c}{March 4-9, 1992}\\
$B$& -22.157\,$\pm$\,0.041 & -0.067\,$\pm$\,0.013 & 0.294\,$\pm$\,0.011\\
$V$& -22.697\,$\pm$\,0.019 & -0.033\,$\pm$\,0.005 & 0.198\,$\pm$\,0.005\\
$R$& -22.768\,$\pm$\,0.036 & -0.002\,$\pm$\,0.018 & 0.170\,$\pm$\,0.010\\
$I$& -22.063\,$\pm$\,0.038 & -0.008\,$\pm$\,0.027 & 0.118\,$\pm$\,0.012\\
\hline
\hline
\end{tabular}
\label{mag0tab}
\end{center}
\end{table}
}


%
\begin{table}
\begin{center}
\caption[]{
Calibration coefficients determined for the different observing runs on 
the UKIRT.
}
\begin{tabular}{ccc}
\ \\
\hline
\hline
\ \ \ \ color & zero point ($c_0$) & extinction coefficient ($c_2$)\\
\hline
\multicolumn{3}{c}{September 28-30, 1991}\\
$H$& -20.500$\pm$0.200 & ---\\
$K^{\prime}$& -20.018$\pm$0.040 & 0.087$\pm$0.032\\
\hline
\multicolumn{3}{c}{February 20-22, 1992}\\
$H$& -20.704$\pm$0.032 & 0.147$\pm$0.048\\
$K$& -20.497$\pm$0.032 & 0.119$\pm$0.047\\
\hline
\hline
\end{tabular}
\vspace{-3ex}
\label{mag0tabir}
\end{center}
\end{table}

\end{document}